\documentclass[aps,twocolumn]{revtex4}
\usepackage{epsf}
\usepackage{dcolumn}
\usepackage{longtable}
\usepackage{graphicx}

\begin{document}

\title{
A Novel Symmetric Four Dimensional Polytope Found Using Optimization Strategies Inspired by Thomson's Problem of Charges on a Sphere}

\author{ Eric Lewin Altschuler$^1$, Antonio P\'erez--Garrido$^2$ and Richard Stong$^3$}

\affiliation{
$^1$Department of Physical Medicine and Rehabilitation.
UMDNJ\\
 \hbox{30 Bergen St., ADMC 1, Suite 101 Newark, NJ 07101, USA}\\
email: eric.altschuler@umdnj.edu\\
 $^2$Departamento de F\'\i sica Aplicada, UPCT\\
\hbox{Campus Muralla del Mar, Cartagena, 30202 Murcia, Spain}\\
email:Antonio.Perez@upct.es\\
$^3$Department of Mathematics, Rice University\\
 Houston, Texas, 77005 USA}

\begin{abstract}
Inspired by, and using methods of optimization derived from classical three dimensional electrostatics, we note a novel beautiful symmetric four dimensional polytope we have found with 80 vertices.  We also describe how the method used to find this symmetric polytope, and related methods can potentially be used to find good examples for the kissing and packing problems in $D$ dimensions.

\end{abstract}

\maketitle

Using an optimization method (described below) inspired by ones ourselves and others have used (see \cite{AP05, EH95, EH91,EH97,AP06}, and refs.\ therein) for a problem in three dimensional electrostatics--Thomson's \cite{Th04} problem of finding the minimum energy of $N$ unit point charges on the surface of a unit conducting sphere--we have found a novel 
beautiful symmetric configuration with 80 vertices (Fig.\ 1).  
The polytope has 64 vertices with 12 nearest neighbors of 0.7624, 0.6707, 0.7654  and 0.6661 distances, 16 vertices with 10 neighbors of   0.7654  and 0.6661 distances.  We have not seen this polytope previously \cite{Co40,Ol06}

\begin{figure}
\begin{center}
\leavevmode
\includegraphics[angle=-90,width=7.5cm]{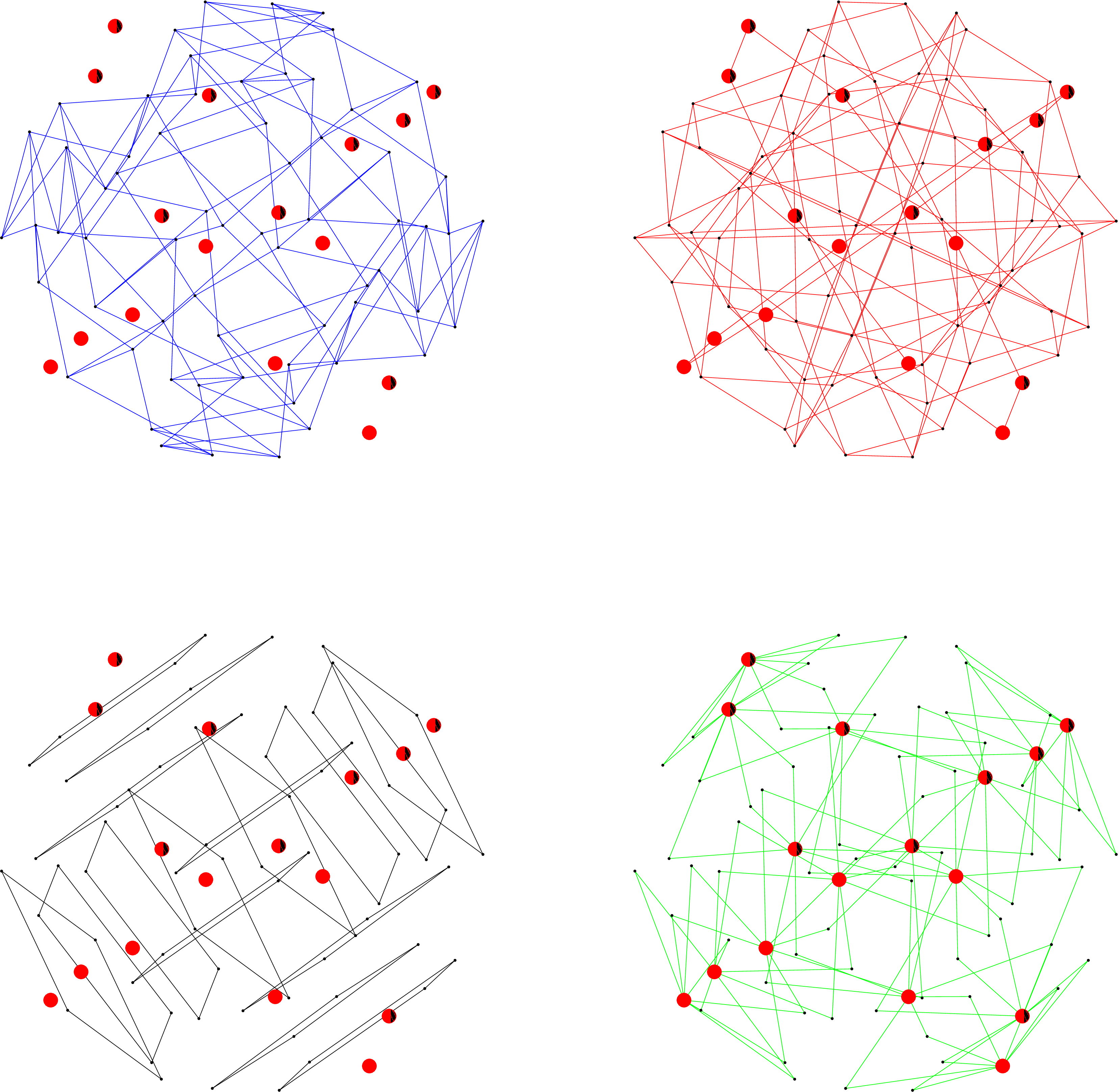}
\end{center}
\caption{
A novel symmetric four dimensional polytope with 80 vertices. Four views showing each a different type of
nearest neighbor bond (different distance). Red dots correspond to a vertex with 10 nearest neighbors.
These figures show a simple parallel projection from 4D to 2D, i.e. $z$ and $w$ coordinates are discarded, 
then each point is  plotted accordingly to its $x$ and $y$ coordinates.  All of the views 
have been rotated to show a symmetric pattern in 2D.
}
\end{figure}

We found the configuration looking at the slightly artificial, in four dimensions, but potentially useful (see below), problem of finding the minimum energy configuration of $N$ {\it charges} (points) on the surface of the hypersphere ($S_3$) $ x^2 + y^2 + z^2 + w^2 =1$ in four dimensions with the {\it energy} function $\sum_{i\neq j} 1/r_{ij}$ where $r$ is the Euclidean distance between two points 1 and 2 $\sqrt{(x_1 - x_2)^2 +(y_1 - y_2)^2 +(z_1 -z_2)^2 +(w_1 - w_2)^2 }$. 
 As most surely like in three dimensions \cite{EH95}, in four or more dimensions the number of good local mimima for this problem grows exponentially with $N$,  and thus we cannot be certain that for $N = 80$ or other $N$ that we have found the global minima. Nevertheless, even good local minima may be interesting or important configurations.   In this initial work we have found our best local minimum for a given $N$ by starting from 100 random initial starting configurations and then used a standard conjugate gradient optimization. We have looked at $N$ = 2 to 200.  The other $N$s for which we have thus far found  nice symmetric configurations are for $N = 5, 8, 24$ and $120$ for which we found as our minimum energy configuration the  simplex 
(4D equivalent of the  tetrahedron), the 16 cell (4D equivalent of the octahedron), the 24 cell and the 600 cell 
(4D equivalent of the icosahedron), respectively  (Fig. 2), four of the six completely regular Platonic solids in four dimensions. We did not find the other two regular polytopes, i.e.  $N=16$ the tesseract (or hypercube, 4D equivalent of the cube) and $N=600$ the 120 cell (4D equivalent of the dodecahedron). Their geometries are not energy minima, similar to what happens with the cube and the dodecahedron in 3D Thomson's problem (Ref.\ \cite{EH97} and references therein).
Using a method related to ours other higher dimensional polytopes have been found\cite{Ho06}.

\begin{figure}
\begin{center}
\leavevmode
\includegraphics[angle=-90,width=7.5cm]{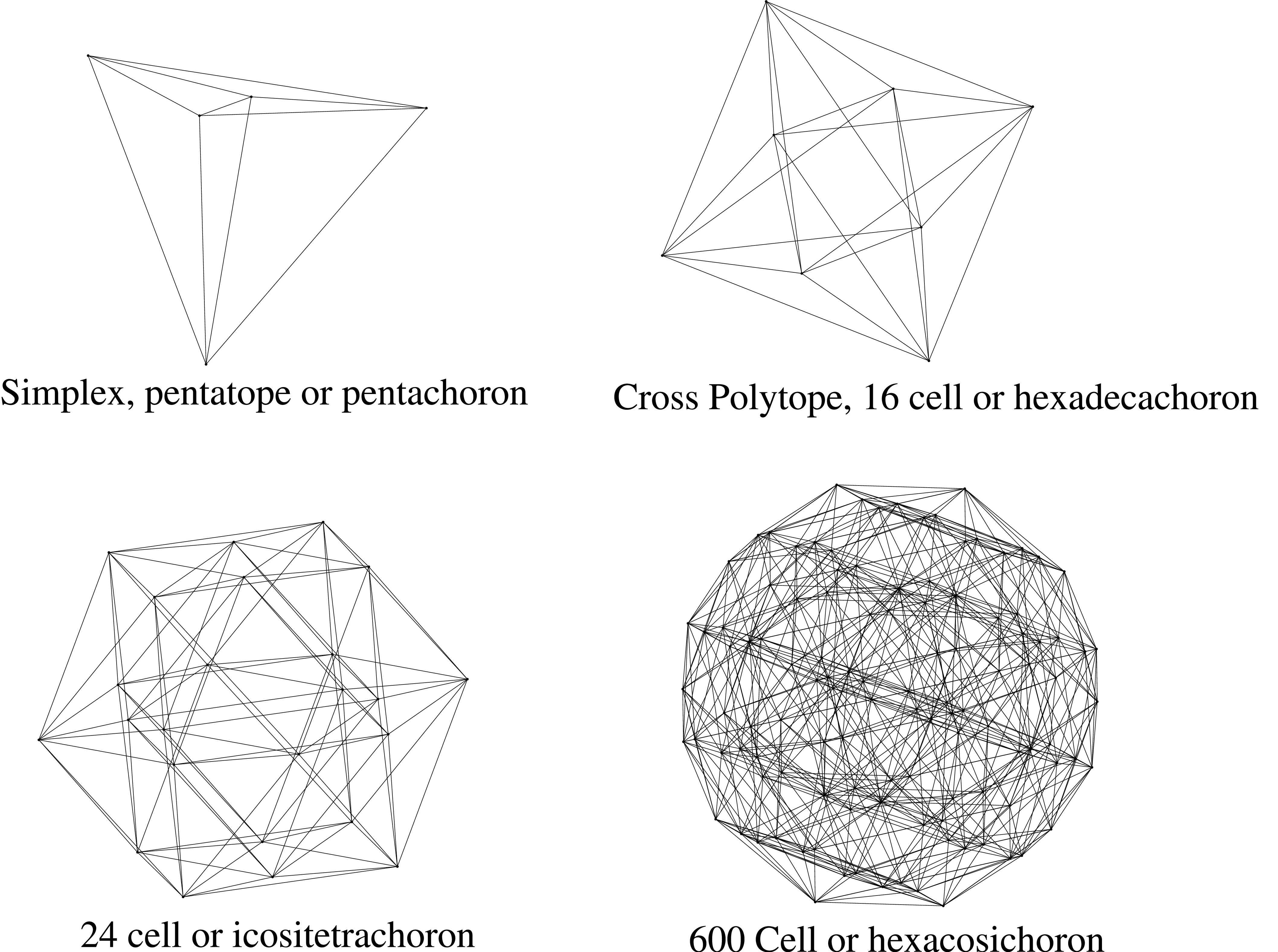}
\end{center}
\caption{
 Regular polytopes found with Thomson's problem minimization process. These figures 
 are  simple parallel projections from 4D to 2D.
}
\end{figure}

The $D$ dimensional kissing problem asks the number of non--intersecting $D$ dimensional unit hyperspheres that can touch a central unit hypersphere of dimension $D$.  The maximal number of such spheres is known as the kissing number for that dimension, $K_D$.  The kissing number is only proven in a few dimensions:
 $K_1$=2, $K_2=6$, $K_3=12$, $K_8=240$, $K_{24}=196569$\cite{Sl98}.
 ($K_4=24$ seems also to have been proven \cite{Mu06}).  See \cite{Sl98} and refs.\ therein for the best known kissing 
number in other dimensions.  The approach of minimizing $\sum_{i\neq j} 1/r_{ij}$ in $D$ dimensions potentially may be useful in finding new larger kissing numbers.  Indeed, if for a given $D$ with largest known kissing number $K_D$ it can be found that a minimum energy configuration with $K_D + 1$ (or more) charges/points can be found with each pair of points at least $D$--dimensional Euclidean distance one from each other, then a new kissing number for that dimension is found.  There is freedom in choosing the energy function to be minimized, e.g. taking $E = \sum_{i\neq j} 1/r_{ij}^n$  where $n$ is an integer $> 1$.  Also we note that for a 
configuration of $N$ points found in $D$ dimensions with greatest separation $s$ between pairs of points, one thus immediately has 
an analogue of a kissing problem of putting hyperspheres of radius at least $s$ around a central unit hypersphere.

Now, a hypersphere has positive curvature and thus is not completely the same as flat space.  To use minimization of electrostatic problems to study the $D$--dimensional packing problem--to find the highest density of $D$--dimensional hyperspheres that can be packed into an infinite D-dimensional (flat) space--an analogue of the above procedure
can  be used by working on the $D$ dimensional surface of a  $D+1$ dimensional torus: 
Indeed, for example, the smallest distance between points on the two dimensional surface of the standard, simple three dimensional torus (doughnut/bagel) determines the density to which two dimensional spheres/balls--i.e., disks--can be packed.
 
The clearest parameterization of tori we have found for this situation is to
describe the $D$ dimensional surface of a ($D+1$ dimensional) torus by $D$ coordinates
($x_1, ..., x_D$) where each $x_i$ lies in [0, 1).  (So each $x_i$ is an angle scaled to lie in [0, 1).)  For any of the $x_i$ coordinates define $\left|\left|x_i - y_i\right|\right| =\left|x_i - y_i\right|$ if $\left|x_i-y_i\right| \leq 1/2$
or $1- \left|x_i - y_i\right|$ if $\left|x_i - y_i\right| > 1/2$.  Then the appropriate squared intrinsic distance between two points on the surface of the torus is  $\left|\left|(x_1,x_2,...,x_n) - (y_1,y_2,...,y_n)\right|\right|^2 = \sum_i \left|\left|x_i-y_i\right|\right|^2$.  One then seeks configurations of points/unit charges that
minimize the energy function $E = \sum_{i\neq j} 1/\left|\left|x-y\right|\right|^n$ where $n$ is an integer $\geq 1$.
After the minimization is complete define $r =  1/2 \cdot min \left|\left|x-y\right|\right|$, where the minimum is taken over all pairs of points.  The volume of the torus is simply $1^D = 1$.
The volume of the $D$ dimensional hyperspheres which can be packed on the surface of the torus 
(and also the packing density/packing fraction since the torus volume equals 1)
is $N\cdot\Omega_D\cdot r^D$, where $\Omega_D = \pi^{D/2}\cdot \Gamma\left((D+2)/2\right)$ and $N$ is the number of points on the surface of the torus.    
From the packing of points on the $D$ dimensional surface of the $D+1$ dimensional torus one then immediately gets a packing for infinite $D$ dimensional flat space by placing a hypersphere of radius r at every point in $R^D$ which differs from the coordinates of a point on the torus by integer amounts in each coordinate.

We thank Andrew M. Gleason for helpful discussions. A.P.G. would like to acknowledge  financial support from Spanish MCyT under grant No. MAT2003--04887.

\end{document}